\documentclass{symmetry10}
\begin{document}

\FirstPageHeading{Vaneeva}

\ShortArticleName{Reduction operators of 
diffusion equations with an exponential source} 

\ArticleName{Reduction operators of variable coefficient\\ semilinear
diffusion equations with\\ an exponential source}

\Author{O.O. VANEEVA~$^\dag$, R.O.~POPOVYCH~$^{\dag\ddag}$ and C. SOPHOCLEOUS~$^\S$}
\AuthorNameForHeading{O.O. Vaneeva, R.O. Popovych and C. Sophocleous}
\AuthorNameForContents{VANEEVA O.O., POPOVYCH R.O. and SOPHOCLEOUS C.}
\ArticleNameForContents{Reduction\\ operators of variable coefficient semilinear
diffusion equations with\\ a power source}

\Address{$^\dag$~Institute of Mathematics of NAS of Ukraine, \\
$\phantom{{}^\dag}$\ 3 Tereshchenkivska Str., 01601 Kyiv-4, Ukraine}
\EmailD{vaneeva@imath.kiev.ua, rop@imath.kiev.ua}

\Address{$^\ddag$~Fakult\"at f\"ur Mathematik, Universit\"at Wien, \\
$\phantom{{}^\ddag}$\ Nordbergstra{\ss}e 15, A-1090 Wien, Austria}

\Address{$^\S$~Department of Mathematics and Statistics,\\
$\phantom{{}^\S}$\ University of Cyprus, Nicosia CY 1678, Cyprus}
\EmailD{christod@ucy.ac.cy}

\Abstract{
Reduction operators (called also nonclassical or $Q$-conditional symmetries) 
of variable coefficient semilinear reaction-diffusion equations 
with exponential source $f(x)u_t=(g(x)u_x)_x+h(x)e^{mu}$ are investigated using the algorithm 
involving a mapping between classes of differential equations, which is generated by a family of point transformations. 
A special attention is paid for checking whether reduction operators are inequivalent to Lie symmetry operators. 
The derived reduction operators are applied to construction of exact solutions.
}

\vspace{-2ex}

\section{Introduction}

Various processes in nature are successfully modeled by nonlinear systems of partial differential equations (PDEs).
In order to study behavior of these processes, it is important to know
exact solutions of corresponding model equations.
Lie symmetries and the classical reduction method present a powerful and algorithmic technique for
the construction of exact solutions (of systems) of PDEs~\cite{vaneeva:Olver1986,vaneeva:Ovsiannikov1982}. 
In~\cite{vaneeva:Bluman&Cole1969} Bluman and Cole introduced a new method for finding exact solutions of PDEs.
It was called ``non-classical'' to emphasize its difference from the classical Lie reduction method.
A precise and rigorous definition of nonclassical invariance was firstly formulated in~\cite{vaneeva:Fushchych&Tsyfra1987} 
as ``a generalization of the Lie definition of invariance'' (see also~\cite{vaneeva:Zhdanov&Tsyfra&Popovych1999}).
Later, operators satisfying the nonclassical invariance criterion were called, by different authors, 
nonclassical symmetries, conditional symmetries and $Q$-conditional symmetries 
\cite{vaneeva:Fushchych&Shtelen&Serov1989,
vaneeva:Fushchych&Shtelen&Serov&Popovych1992,
vaneeva:Fushchych&Zhdanov1992,vaneeva:Levi&Winternitz1989}.
Until now all names are in use. 
See, e.g., \cite{vaneeva:Kunzinger&Popovych2009,
vaneeva:Olver&Vorob'ev1996,vaneeva:Saccomandi2004} for comprehensive reviews of the subject.  
Following Ref.~\cite{vaneeva:Popovych&Vaneeva&Ivanova2007} we call nonclassical symmetries \emph{reduction operators}. 
The necessary definitions, including ones of equivalence of reduction operators, and statements relevant for this paper 
are collected in~\cite{vaneeva:VPS_2009}.

The problem of finding reduction operators for a PDE 
reduces to integration of an overdetermined system of nonlinear PDEs. 
The complexity increases in times in the case of classification problem of reduction operators for
a class of PDEs having nonconstant arbitrary elements.

The experience of  classification of Lie symmetries for classes of variable coefficient
PDEs shows that the usage of equivalence and gauging
transformations can essentially simplify the group classification problem and even be a crucial point in solving 
the 
problem~\cite{vaneeva:Ivanova&Popovych&Sophocleous2006Part1,
vaneeva:VJPS2007,vaneeva:VPS_2009}.
The above transformations are of major importance for studying reduction operators
since under their classification one needs to overcome  much more essential obstacles
then those arising under the classification of Lie symmetries.

In~\cite{vaneeva:VPS_2009} we propose  an algorithm involving mapping between classes
for finding reduction operators of the variable coefficient reaction-diffusion equations with power nonlinearity
\begin{equation}\label{vaneeva:eqRDfghPower}
f(x)u_t=(g(x)u_x)_x+h(x)u^m,
\end{equation}
where $f$, $g$ and $h$  are arbitrary smooth functions of the variable~$x$ and 
$m$ is an arbitrary constant such that $fgh\neq0$ and $m\neq0,1$.
In~\cite{vaneeva:VPS_Cyprus_proc08}  reduction operators of 
the equations from class~\eqref{vaneeva:eqRDfghPower} with $m\neq2$ were investigated using this algorithm.
The case $m=2$ was not systematically considered since it is singular from the Lie symmetry point of view and needs
an additional mapping between classes (see~\cite{vaneeva:VPS_2009} for more details).
Nevertheless, all the reduction operators constructed  
in~\cite{vaneeva:VPS_Cyprus_proc08} for the general case $m\ne0,1,2$
are also fit for the values $m=0,1,2$.

In this paper we implement the same technique to find reduction operators of
the variable coefficient reaction-diffusion equations with exponential nonlinearity
\begin{equation}\label{vaneeva:eqRDfghExp_n=0}
f(x)u_t=(g(x)u_x)_x+h(x)e^{mu}.
\end{equation}
Here $f$, $g$ and $h$  are arbitrary smooth functions of the variable~$x$,
$fgh\neq0$ and $m$ is an arbitrary nonvanishing constant.

The structure of this paper is as follows. For convenience of readers
section~\ref{vaneeva:Sect_Lie_sym} contains a short review of results obtained 
in~\cite{vaneeva:vaneeva_exp} and used here.
Namely, in this section the necessary information concerning equivalence transformations, 
the mapping of class~\eqref{vaneeva:eqRDfghExp_n=0} to the so-called ``imaged'' class and 
the group classification of equations from the imaged class is collected. 
Moreover, all additional equivalence transformations connecting the cases of Lie symmetry extensions (cf.\ table~1) 
are first found and presented therein. 
As a result, the classifications of Lie symmetry extensions up to all admissible point transformations in the imaged and, therefore, initial classes 
are also obtained.
Section~\ref{vaneeva:Sect_Algorithm} is devoted to the description of  the original algorithm for finding reduction operators of equations from class~\eqref{vaneeva:eqRDfghExp_n=0} 
using a mapping between classes generated by a family of point transformations.
The results of section~\ref{vaneeva:Sect_RedOpsAndExactSolutions} are completely new and concern the investigation 
of reduction operators for equations from the imaged class. 
Reduction operators obtained in an explicit form is used for the construction of exact solutions of equations from both the imaged and initial classes.

\section{Lie symmetries and equivalence transformations}\label{vaneeva:Sect_Lie_sym}

Class~\eqref{vaneeva:eqRDfghExp_n=0} has complicated transformational properties. 
An indicator of this is that it possesses the nontrivial generalized extended equivalence group, 
which does not coincide with its usual equivalence group, cf.\ Theorem~\ref{vaneeva:TheoremOnGenExtEquivGroupOfeqRDfghExp_n=0} below.
To produce group classification of class~\eqref{vaneeva:eqRDfghExp_n=0},
it is necessary to gauge arbitrary elements of this class with equivalence transformations and
a subsequent mapping of it onto a simpler 
class~\cite{vaneeva:VPS_2009,vaneeva:vaneeva_exp}. 
It appears that the preimage set of each equation from the imaged class is a two-parametric family of equations from the initial class~\eqref{vaneeva:eqRDfghExp_n=0}. 
Moreover, preimages of the same equation belong to the same orbit of the equivalence group of the initial class. It allows
one to look only for simplest representative of the preimage to obtain its symmetries, exact solution, etc., 
and then reproduce these results for two-parametric family of equations from the initial class using equivalence transformations.

\begin{theorem}\label{vaneeva:TheoremOnGenExtEquivGroupOfeqRDfghExp_n=0} 
The generalized extended equivalence group~$\hat G^{\sim}_{\rm exp}$ of class~\eqref{vaneeva:eqRDfghExp_n=0}  consists of the transformations
\[
\begin{array}{l}
\tilde t=\delta_1 t+\delta_2,\quad \tilde x=\varphi(x),\quad
\tilde u=\delta_3u+\psi(x), \\[1ex]
\tilde f=\dfrac{\delta_0\delta_1}{\varphi_x}f, \quad
\tilde g=\delta_0\varphi_x g,\quad
\tilde h=\dfrac{\delta_0\delta_3}{\varphi_x}e^{-\frac{m\psi(x)}{\delta_3}}h,
\quad \tilde m=\dfrac{m}{\delta_3},
\end{array}
\]
where $\varphi$ is an arbitrary non-constant smooth function of~$x$, $\psi=\delta_4\int\frac{dx}{g(x)}+\delta_5$,
and $\delta_j,$ $j=0,1,\dots,5,$ are arbitrary constants such that $\delta_0\delta_1\delta_3\not=0$.
\end{theorem} 

\begin{corollary}
The usual equivalence group of class~\eqref{vaneeva:eqRDfghExp_n=0} is the subgroup of~$\hat G^{\sim}_{\rm exp}$ singled out by the condition $\delta_4=0$. 
\end{corollary}

The transformations from~$\hat G^{\sim}_{\rm exp}$ associated with varying the parameter~$\delta_0$ in fact do not change equations from class~\eqref{vaneeva:eqRDfghExp_n=0} 
and hence form the gauge equivalence group of this class. 
The values of arbitrary elements connected by a such transformation correspond to different representations of the same equation. 

Analogously to the power case we at first map class~\eqref{vaneeva:eqRDfghExp_n=0} onto its subclass 
\begin{gather}\label{vaneeva:class_f=g_exp}
f(x)u_{t}= (f(x)u_x)_x+h(x)e^u
\end{gather}
(we omit tildes over the variables), using the family of equivalence transformations 
parameterized by the arbitrary elements~$f$, $g$ and~$m$,
\begin{equation}\label{vaneeva:gauge_f=g_exp}
\tilde t={\rm sign }(f(x)g(x))\,t,\quad 
\smash{\tilde x=\int\left|\frac{f(x)}{g(x)}\right|^\frac12dx,}\vphantom{\frac bb} \quad
\tilde u=m\,u.
\end{equation}
The new arbitrary elements are expressed via the old ones in the following way:  
\[
\tilde f(\tilde x)=\tilde g(\tilde x)={\rm sign}(g(x))|f(x)g(x)|^\frac12,\quad  
\smash{\tilde h(\tilde x)=m\left|\frac{g(x)}{f(x)}\right|^\frac12h(x),}\vphantom{\frac bb} \quad 
\tilde m=1.
\]

The next step is to change the dependent variable in class~\eqref{vaneeva:class_f=g_exp}:
\begin{equation}\label{vaneeva:gauge_exp}
v(t,x)=u(t,x)+G(x),\quad\mbox{where}\quad G(x)=\ln|f(x)^{-1}h(x)|.
\end{equation}
Finally we obtain the class
\begin{equation}\label{vaneeva:class_vFH_exp}
v_t=v_{xx}+F(x)v_x+\varepsilon e^{\,v}+H(x),
\end{equation}
where $\varepsilon={\rm sign}(f(x)h(x))$ and the new arbitrary elements $F$ and $H$ are expressed 
via the arbitrary elements of class~\eqref{vaneeva:class_vFH_exp} according to the formulas
\begin{equation}\label{vaneeva:eqfF}
F=f_xf^{-1},\quad H=-G_{xx}-G_xF.
\end{equation}

All results on Lie symmetries and exact solutions of class~\eqref{vaneeva:class_vFH_exp}
can be extended to class~\eqref{vaneeva:class_f=g_exp} by the inversion of transformation~\eqref{vaneeva:gauge_exp}.

The arbitrary elements $f$ and $h$ of class~\eqref{vaneeva:class_vFH_exp} are expressed via the functions $F$ and $H$ in the following way:
\begin{gather}\label{vaneeva:Eq_fh_exp}\arraycolsep=0ex
\begin{array}{l}
f=c_0\exp\left(\int Fdx\right),\quad h=\varepsilon c_0\exp\left(\int Fdx+G\right),\\[1ex]
\mbox{where}\quad G=\int e^{-\int Fdx}\left(c_1-\int He^{\int Fdx} dx\right)dx+c_2.
\end{array}\end{gather}
Here $c_0$, $c_1$ and $c_2$ are arbitrary constants, $c_0\neq0$. 
The constant $c_0$ is inessential and can be set to the unity by an obvious gauge equivalence transformation.
The equations from class~\eqref{vaneeva:class_f=g_exp}, which have 
the same image in class~\eqref{vaneeva:class_vFH_exp} with respect to transformation~\eqref{vaneeva:gauge_exp}, 
i.e., whose arbitrary elements are given by~\eqref{vaneeva:Eq_fh_exp} and differ only by values of constants $c_1$ and $c_2$,  
are $\hat G^{\sim}_{\rm exp}$-equivalent.
The equivalence transformation 
\begin{equation}\label{vaneeva:eq_equiv_tr_exp}\textstyle
\tilde t=t,\quad\tilde x=x,\quad\tilde u=u+c_1\int e^{-\int Fdx}dx+c_2
\end{equation}
maps an equation~\eqref{vaneeva:class_vFH_exp} having $f$ and $h$ of the form~\eqref{vaneeva:Eq_fh_exp} with $c_1^2+c_2^2\neq0$ to the one
with $c_1=c_2=0$. Hence up to $\hat G^{\sim}_{\rm exp}$-equivalence 
we can consider, without loss of generality, only equations from
class~\eqref{vaneeva:class_f=g_exp} that have the arbitrary elements determined by~\eqref{vaneeva:Eq_fh_exp} with $c_1=c_2=0$.

\begin{theorem}
The generalized extended equivalence group~$G^{\sim}_{\rm exp}$ of class~\eqref{vaneeva:class_vFH_exp} coicides with its usual equivalence group 
and is formed by the transformations
\[\begin{array}{l}
\tilde t=\delta_1^{\,2} t+\delta_2,\quad \tilde x=\delta_1x+\delta_3, \quad
\tilde v=v-\ln \delta_1^{\,2},\\[1ex]
\tilde F=\delta_1^{\,-1}F, \quad
\tilde H=\delta_1^{\,-2}H,
\end{array}\]
where $\delta_j,$ $j=1,2,3,$  are arbitrary constants, $\delta_1\not=0$.
\end{theorem}

The \emph{kernel} of the maximal Lie invariance algebras of equations from class~\eqref{vaneeva:class_vFH_exp}
is the one-dimensional algebra $\langle\partial_t\rangle$. 
It means that any equation from class~\eqref{vaneeva:class_vFH_exp} is invariant with respect to translations by $t$, 
and there are no more common Lie symmetries. 

\begin{theorem}
$G^{\sim}_{\rm exp}$-inequivalent cases of extension of the maximal Lie invariance 
algebras in class~\eqref{vaneeva:class_vFH_exp} are exhausted by ones
presented in table~\ref{vaneeva:TableNonclassicalSym_exp}.
\end{theorem}

\begin{center}\renewcommand{\arraystretch}{1.8}
\refstepcounter{table}\label{vaneeva:TableNonclassicalSym_exp}\textbf{Table~\thetable.} The group classification of the class~$v_t=v_{xx}+F(x)v_x+\varepsilon e^{\,v}+H(x).$
\\[2ex]
\begin{tabular}{|c|c|c|c|l|}
\hline
N&$F(x)$&$H(x)$&\hfil Basis of $A^{\rm max}$ \\
\hline
0&$\forall$&$\forall$&$\partial_t$\\
\hline
1&$\alpha x^{-1}+\mu x$&$\beta x^{-2}+2\mu $&$\partial_t,\,e^{-2\mu t}(\partial_t-\mu x\partial_x+2\mu \partial_v)$\\
\hline
2&$\alpha x^{-1}$&$\beta x^{-2}$&$\partial_t,\,2t\partial_t+x\partial_x-2\partial_v$\\
\hline
3&$\mu x$&$\gamma$&$\partial_t,\,e^{-\mu t}\partial_x$\\
\hline

4&$\lambda$&$\gamma$&$\partial_t,\,\partial_x$\\
\hline
5&$\mu x$&$2\mu $&$\partial_t,\,e^{-\mu t}\partial_x,\,e^{-2\mu t}(\partial_t-\mu x\partial_x+2\mu \partial_v)$\\
\hline
6&$\lambda$&$0$&$\partial_t,\,\partial_x,\,2t\partial_t+(x-\lambda t)\partial_x-2\partial_v$\\

\hline
\end{tabular}
\\[2ex]
\parbox{135mm}{Here   $\lambda\in\{0,1\}\bmod G^{\sim}_{\rm exp}$, $\mu=\pm1\bmod G^{\sim}_{\rm exp}$; $\alpha, \beta, \gamma$ are arbitrary constants,
$\alpha^2+\beta^2\neq0$.
We also have $\gamma\neq2\mu$ and $\gamma\ne0$ in cases 3 and 4, respectively.}
\end{center}

The corresponding results on group classification of class~\eqref{vaneeva:eqRDfghExp_n=0} up to $\hat G^{\sim}_{\rm exp}$-equiv\-alence
is given in table 3 of~\cite{vaneeva:vaneeva_exp}.

Additional equivalence transformations between $G^{\sim}_{\rm exp}$-inequivalent cases of Lie symmetry extension are also constructed.
The pairs of point-equivalent cases from table~\ref{vaneeva:TableNonclassicalSym_exp} and the corresponding transformations
are exhausted by the following:
\begin{gather}\label{vaneeva:eq_add_eq_tr_exp}\begin{split}
&1\mapsto{\tilde 2},\quad 5\mapsto{\tilde6}|_{\tilde\lambda=0}\colon\,\,
\tilde t=\frac 1{2\mu}e^{2\mu t},\quad\tilde x=e^{\mu t}x,\quad \tilde v=v-2\mu t,\\
&4\mapsto{\tilde4}|_{\tilde\lambda=0},\quad6\mapsto{\tilde6}|_{\tilde\lambda=0}\colon\,\,\tilde t=t,\quad \tilde x=x+\lambda t,\quad \tilde v=v.
\vphantom{\frac 1{2\mu}}
\end{split}\end{gather}
The inequivalence of other different cases of table~1 can be proved using differences 
in properties of the corresponding maximal Lie invariance algebras, which should coincide for similar equations. 
Thus, the dimensions of the maximal Lie invariance algebras are one, three and two 
in the general case (case~0), cases~5 and~6 and the other cases, respectively. 
In contrast to cases 1--3, the algebra of case~4 is commutative. 
The derivative of the algebra of case~3 has the zero projection on the space of~$t$, and this is not the case for cases~1 and~2.
Possessing the zero (resp.\ non-zero) projection on the space of~$t$ is 
an invariant characteristic of Lie algebras of vector fields in the space of the variables~$t$, $x$ and~$v$
with respect to point transformations connecting a pair of evolution equations 
since for any such transformation the expression of the transformed $t$ is well known to depend only on~$t$ 
\cite{vaneeva:Kingston&Sophocleous1998,vaneeva:Magadeev1993}.

More difficult problem is to prove that there are no more additional equivalences within a parameterized case of table~1. 
(In fact, all the cases are parameterized.) 
This needs at least a preliminary study of 
form-preserving~\cite{vaneeva:Kingston&Sophocleous1998} 
(or admissible~\cite{vaneeva:Popovych2006c,
vaneeva:Popovych&Kunzinger&Eshraghi2010}) transformations. 
In contrast to transformations from the corresponding equivalence group, 
which transform each equation from the class~$\mathcal L$ of differential equations under consideration to 
an equation from the same class, a form-preserving transformation should transform at least a single equation from~$\mathcal L$ 
to an equation from the same class. 
The notion of admissible transformations is a formalization of the notion of form-preserving transformations. 
The set of admissible transformations of the class~$\mathcal L$ is formed by the triples 
each of which consists of the tuples of arbitrary elements corresponding to the initial and target equations 
and a point transformation connecting these equations. 
It is obvious that each transformation from the equivalence group generates a family of admissible transformations 
parameterized by arbitrary elements of the class~$\mathcal L$. 

A preliminary description of the set of admissible transformations of the class~\eqref{vaneeva:class_vFH_exp}, 
which is sufficient for our purpose, is given by the following statements. 

\begin{proposition}
Any admissible point transformation in the class~\eqref{vaneeva:class_vFH_exp} has the form
\[
\tilde t=T(t),\quad \tilde x=\delta\sqrt{T_t}\,x+X(t),\quad \tilde v=v-\ln T_t,
\] 
where $\delta=\pm1$ and $T$ and~$X$ are arbitrary smooth functions of~$t$ such that $T_t>0$. 
The corresponding values of the arbitrary elements are related via the formulas
\[
\tilde F=\frac\delta{\sqrt{T_t}}F-\frac{\delta}2\frac{T_{tt}}{\sqrt{T_t{}^3}}x-\frac{X_t}{T_t},\quad 
\tilde H=\frac1{T_t}H-\frac{T_{tt}}{T_t{}^2}.
\]
\end{proposition}

\begin{corollary}
Only equations from the class~\eqref{vaneeva:class_vFH_exp} whose arbitrary elements have the form
\begin{equation}\label{vaneeva:EqClass_vFH_exp_CondForNontrivAdmTrans}
F=\mu x+\lambda+\frac{\alpha}{x+\kappa},\quad 
H=\gamma+\frac{\beta}{(x+\kappa)^2},
\end{equation}
where $\alpha$, $\beta$, $\gamma$, $\kappa$ and $\mu$ are constants,
possess admissible transformations that are not generated by transformations from the equivalence group~$G^{\sim}_{\rm exp}$.
The subclass of the class~\eqref{vaneeva:class_vFH_exp}, singled out by the condition~\eqref{vaneeva:EqClass_vFH_exp_CondForNontrivAdmTrans}, 
is closed under any admissible transformation within the class~\eqref{vaneeva:class_vFH_exp}. 
The (constant) parameters of the representation~\eqref{vaneeva:EqClass_vFH_exp_CondForNontrivAdmTrans} 
are transformed by an admissible transformation in the following way:
\begin{gather*}
\tilde \alpha=\alpha,\quad \tilde\beta=\beta,\qquad
\tilde \kappa=\delta\sqrt{T_t}\kappa-X\quad\mbox{if}\quad(\alpha,\beta)\neq(0,0),\\
\tilde\gamma=\frac{\gamma}{T_t}-\frac{T_{tt}}{T_t^2},\quad
\tilde\mu=\frac{\mu}{T_t}-\frac12\frac{T_{tt}}{T_t^2},\quad
\tilde\lambda=-\tilde\mu X-\frac{X_t}{T_t}+\frac{\delta\lambda}{\sqrt{T_t}}.
\end{gather*}
In particular, $T_{tt}=0$ if $\gamma\neq2\mu$.
\end{corollary}

Finally, we can formulate the assertion on group classification with respect to the set of admissible transformations. 

\begin{theorem}
Up to point equivalence,  cases of extension of the maximal Lie invariance 
algebras in class~\eqref{vaneeva:class_vFH_exp} are exhausted by cases 0, 2, 3, $4_{\lambda=0}$ and~$6|_{\lambda=0}$ 
of table~\ref{vaneeva:TableNonclassicalSym_exp}.
\end{theorem}

\section{Algorithm of finding reduction operators\\ via mappings between classes}\label{vaneeva:Sect_Algorithm}

At first we adduce the definition of nonclassical symmetries 
\cite{vaneeva:Fushchych&Zhdanov1992,
vaneeva:Popovych&Vaneeva&Ivanova2007,vaneeva:Zhdanov&Tsyfra&Popovych1999}, 
adopting it for the case of one second-order PDE with two independent variables, relevant for this paper. 
Consider a second-order differential equation~$\mathcal L$ of the form~$L(t,x,u_{(2)})=0$
for the unknown function $u$ of the two independent variables~$t$ and~$x$,
where $u_{(2)}=(u,u_t,u_x,u_{tt},u_{tx},u_{xx})$. 
Let $Q$ be a first-order differential operator of the general form
\[
Q=\tau(t,x,u)\partial_t+\xi(t,x,u)\partial_x+\eta(t,x,u)\partial_u, \quad (\tau,\xi)\not=(0,0).
\]

\begin{definition}\label{vaneeva:DefinitionOfCondSym_n0}
The differential equation~$\mathcal L$ is called
\emph{conditionally invariant} with respect to an operator $Q$ if
the relation
\begin{equation}\label{vaneeva:eq_cond_sym_criterion}
Q_{(2)}L(t,x,u_{(2)})\bigl|_{\mathcal L\cap\mathcal{Q}^{(2)}}=0
\end{equation}
holds, which is called the \emph{conditional (or nonclassical) invariance criterion}.
Then $Q$ is called \emph{conditional symmetry}
(or nonclassical symmetry, $Q$-conditional symmetry or reduction operator)
of the equation~$\mathcal L$.
\end{definition}

The symbol $Q_{(2)}$ stands for the standard second prolongation of~$Q$ 
(see e.g.~\cite{vaneeva:Olver1986,vaneeva:Ovsiannikov1982}).
$\mathcal Q^{(2)}$~is the manifold determined in the second-order jet space by the differential
consequences of the characteristic equation 
$Q[u]:=\eta-\tau u_t-\xi u_x=0,$
which have, as differential equations, orders not greater than two.

It was proved in~\cite{vaneeva:Zhdanov&Tsyfra&Popovych1999} that a differential equation~$\mathcal L$ is conditionally invariant 
with respect to the operator $Q$ if and only if the ansatz constructed with this operator reduces the equation~$\mathcal L$.
That is why it seems natural to call operators of conditional (nonclassical symmetries) {\it reduction operators}. 

Here we present the algorithm of application of equivalence transformations, gauging of arbitrary elements
and mappings between classes of equations to classification of reduction operators of class~\eqref{vaneeva:eqRDfghExp_n=0}

\begin{enumerate}\itemsep=0ex
\item  
At first we gauge class~\eqref{vaneeva:eqRDfghExp_n=0}
to subclass~\eqref{vaneeva:class_f=g_exp} constrained by the condition~$f=g$.
Then class~\eqref{vaneeva:class_f=g_exp} is mapped to the imaged class~\eqref{vaneeva:class_vFH_exp}
by transformation~\eqref{vaneeva:gauge_exp}.
\item 
Reduction operators should be classified up to the equivalence  relations
generated by the corresponding equivalence groups or even by the whole sets of admissible transformations.
As the singular case $\tau=0$ is 
``no-go''~\cite{vaneeva:Kunzinger&Popovych2008,vaneeva:Zhdanov&Lahno1998}, 
only the regular case $\tau\ne0$ (reduced to the case $\tau=1$) 
should be considered.
Operators equivalent to Lie symmetry ones should be neglected.
\item 
It is well-known (see e.g. \cite{vaneeva:Arrigo&Hill&Broadbridge1993,
vaneeva:Clarkson&Mansfield1993}) 
that the equations from the imaged class~\eqref{vaneeva:class_vFH_exp} with $F=0$ and $H={\rm const}$ 
and, therefore, all equations similar to them with respect to point transformations 
possess no regular reduction operators that are inequivalent to Lie symmetry operators.
This is why all the above equations should be excluded from the consideration.
\item 
Preimages of the nonclassical symmetries obtained and of equations
admitting them should be found using the inverses of gauging transformations
and the push-forwards by these inverses on the sets of operators.
\end{enumerate}

Reduction operators of equations from class~\eqref{vaneeva:class_f=g_exp} 
are easily found from reduction operators of corresponding equations from~\eqref{vaneeva:class_vFH_exp}
using the formula
\begin{gather}\label{vaneeva:tr_op}
\tilde Q=\tau\partial_t+\xi\partial_x+\left(\eta-\xi G_x\right)\partial_u.
\end{gather} 
Here $\tau$, $\xi$ and $\eta$ respectively are the coefficients of $\partial_t$, $\partial_x$ and $\partial_v$
in a reduction operator of an equation from class~\eqref{vaneeva:class_vFH_exp}.
The function $G$ is defined in~\eqref{vaneeva:Eq_fh_exp}.

In~\cite{vaneeva:VPS_Cyprus_proc08,vaneeva:VPS_2009} we discussed two ways to use mappings between classes of equations 
in the investigation of reduction operators and their usage to finding exact solutions.
The preferable way is based on the implementation of reductions in the imaged class and preimaging of the
obtained exact solutions instead of preimaging the corresponding reduction operators.

\section{Reduction operators and exact solutions}\label{vaneeva:Sect_RedOpsAndExactSolutions}

Following the above algorithm, we look for $G^{\sim}_{\rm exp}$-inequivalent reduction operators with nonvanishing 
coefficient of $\partial_t$ for the equations from the imaged
class~\eqref{vaneeva:class_vFH_exp}. Up~to the usual equivalence of reduction operators we need to consider 
only the operators of the form \[Q=\partial_t+\xi(t,x,v)\partial_x+\eta(t,x,v)\partial_v.\]

Applying conditional invariance criterion~\eqref{vaneeva:eq_cond_sym_criterion} to equation~\eqref{vaneeva:class_vFH_exp}
we obtain a third-degree polynomial of $v_x$ with coefficients depending on $t$, $x$ and~$v$, which has to identically equal zero. 
Splitting it with respect to different powers of $v_x$ results in 
the following determining equations for the coefficients $\xi$ and $\eta$:
\begin{gather}\label{vaneeva:EqDetForRedOpsOfImagedEqExp}\arraycolsep=0ex
\begin{array}{l}
\xi_{vv}=0,\qquad
\eta_{vv}=2(\xi_{xv}-\xi\xi_v-F\xi_v),\\[1ex]
\xi_t-\xi_{xx}+2\xi_x\xi+3\xi_{v}\left(H +\varepsilon e^v\right)+2\eta_{vx}-2\xi_{v}\eta+F\xi_x+\xi F_x=0,\\[1ex]
\eta_t-\eta_{xx}+2\xi_x\eta=\xi H_{x}+F\eta_x+(2\xi_x-\eta_v)H+\varepsilon e^v\left(\eta+2\xi_x-\eta_v\right).
\end{array}\end{gather}

Integration of the first two equations of~\eqref{vaneeva:EqDetForRedOpsOfImagedEqExp} gives us 
the expressions for $\xi$ and $\eta$ with an explicit dependence on~$v$:
\begin{gather}\label{vaneeva:Forms_of_xi_and_eta_Exp}\arraycolsep=0ex
\begin{array}{l}
\xi=av+b,\\[.5ex]
\eta=-\dfrac13a^2v^3+(a_x-ab-aF)v^2+cv+d,
\end{array}\end{gather}
where $a=a(t,x),$ $b=b(t,x),$ $c=c(t,x)$ and $d=d(t,x)$ are smooth functions of~$t$ and~$x$.

Substituting the expressions~\eqref{vaneeva:Forms_of_xi_and_eta_Exp} for $\xi$ and $\eta$ 
into the third and forth equations of~\eqref{vaneeva:EqDetForRedOpsOfImagedEqExp}
and collecting the coefficients of different powers of $v$ in the resulting equations, 
we derive the conditions $a=c=0$, $d=-2b_x$ and two classifying equations, 
which contain both the coefficient $b=b(t,x)$ and the arbitrary elements $F=F(x)$ and $H=H(x)$. 
Summing up the above consideration, we have the following assertion. 

\begin{proposition}
Any regular reduction operator of an equation from the imaged class~\eqref{vaneeva:class_vFH_exp} 
is equivalent to an operator of the form
\begin{equation}\label{vaneeva:EqRedOp_exp}
Q=\partial_t+b\partial_x-2b_x\partial_v,
\end{equation}
where the coefficient $b=b(t,x)$ satisfies the overdetermined system of partial differential equations
\begin{gather}\label{vaneeva:EqDet_exp}\arraycolsep=0ex
\begin{array}{l}
b_t-b_{xx}+2bb_x+Fb_x+b F_x=0,\\[1ex]
bH_x+2b_xH-4bb_{xx}-2(Fb)_{xx}-2Fb_{xx}=0
\end{array}
\end{gather}
with the corresponding values of the arbitrary elements $F=F(x)$ and $H=H(x)$.
\end{proposition}

The second equation of~\eqref{vaneeva:EqDet_exp} can be written in the more compact form 
\[
4(b+F)b_{xx}=2Kb_x+K_xb,
\] 
where $K=H-2F_x$, which is more convenient for the study of compatibility.

Analogously to the power case, we were not able to completely study all the cases of integration of system~\eqref{vaneeva:EqDet_exp} 
depending on values of~$F$ and~$H$. 
This is why we try to solve this system under different additional constraints imposed either on~$b$ or on~$(F,H)$. 

The most interesting results are obtained for the constraint $b_t=0$.
Then $F$ and $H$ are expressed, after a partial integration of~\eqref{vaneeva:EqDet_exp}, via the function $b=b(x)$ 
that leads to the following statement. 

\begin{theorem}\label{vaneeva:theorem_Exp}
For an arbitrary smooth function $b=b(x)$
the equation from class~\eqref{vaneeva:class_vFH_exp} with the arbitrary elements  
\begin{gather}\label{vaneeva:eq_FH_exp}\arraycolsep=0ex
\begin{array}{l}
F=\dfrac1b\left(b_x+k_1-b^2\right),\\[1.5ex]
H=\dfrac2{b^2} \left(k_2+b_x(k_1-b^2) +bb_{xx}\right),
\end{array}\end{gather}
where $k_1$ and $k_2$ are constants, admits the reduction operator~\eqref{vaneeva:EqRedOp_exp} with the same~$b$.
\end{theorem}

An ansatz constructed by the reduction operator~\eqref{vaneeva:EqRedOp_exp} with $b_t=0$ has the form 
\begin{equation*}v=z(\omega)-2\ln |b|,\quad\mbox{where}\quad\omega=t-\int \frac {dx}{b},
\end{equation*} 
The substitution of the ansatz into equation~\eqref{vaneeva:class_vFH_exp} leads to the reduced ODE
\begin{equation}\label{vaneeva:Eq_Reduced_exp}
z_{\omega\omega}-k_1z_\omega+\varepsilon e^{\,z}+2k_2=0.
\end{equation}

For $k_1=0$  the general solution of~\eqref{vaneeva:Eq_Reduced_exp} is written in the implicit form
\begin{gather}\label{vaneeva:eq_implicit_solution_exp}
\int\!\left(c_1-4k_2z-2\varepsilon e^{\,z}\right)^{-\frac12}dz=\pm(\omega+c_2).
\end{gather}
Up to similarity of solutions of equation~\eqref{vaneeva:class_vFH_exp}, the constant~$c_2$ is inessential 
and can be set to equal zero by a translation of~$\omega$, which is always induced by a translation of~$t$. 

Setting additionally $k_2=0$ in~\eqref{vaneeva:eq_implicit_solution_exp}, we are able to integrate in~\eqref{vaneeva:eq_implicit_solution_exp} in a closed form 
and to  explicitly write down the general solution of~\eqref{vaneeva:Eq_Reduced_exp}.
If $\varepsilon=1$ then $c_1>0$ and~\eqref{vaneeva:eq_implicit_solution_exp} gives the following expression for $e^z$:
\[
e^z=\dfrac{2s_1^2}{\cosh^{2}(s_1\omega+s_2)}.
\]
Here and below $s_1=\sqrt{|c_1|}/2$ and $s_2=c_2s_1.$ If $\varepsilon=-1$, the integration leads to 
\begin{gather*}e^z=\begin{cases}\dfrac{2s_1^2}{\sinh^{2}(s_1\omega+s_2)},& c_1>0,\\[2ex]
\dfrac{2s_1^2}{\cos^{2}(s_1\omega+s_2)},& c_1<0,\\[2ex]
\dfrac2{(\omega+c_2)^2},& c_1=0.
\end{cases}\end{gather*}
As a result, for the equation from class~\eqref{vaneeva:class_vFH_exp} of the form
\begin{gather}\label{vaneeva:eq_with_exact_solution_exp}
v_t=v_{xx}+\frac1b\left(b_x-b^2\right)v_x+\varepsilon e^v+\frac2{b} \left(b_{xx}-bb_x\right)
\end{gather} 
with $\varepsilon=-1$   we construct three families of exact solutions 
\begin{gather}\nonumber
v=-2\ln\left|\dfrac{\sqrt 2}{2s_1}\,b\,
{\sinh\left( s_1t- s_1\int\frac{dx}b+s_2\right)}\right|,\\[.5ex]\label{vaneeva:eq_exact_solution_exp1}
v=-2\ln\left|\dfrac{\sqrt 2}{2s_1}\,b\,
{\cos\left( s_1t- s_1\int\frac{dx}b+s_2\right)}\right|,\\[.5ex]\nonumber
v=-2\ln\left|\frac{\sqrt{2}}2\,b\,\left(t-\int\frac{dx}b+c_2\right)\right|,
\end{gather}
where $s_1$, $s_2$ and~$c_2$ are arbitrary constants, $s_1\ne0$. 
Also we obtain a family of exact solution 
\begin{gather}\label{vaneeva:eq_exact_solution_exp2}
v=-2\ln\left|\dfrac{\sqrt 2}{2s_1}\,b\,{\cosh\left( s_1t- s_1\int\frac{dx}b+s_2\right)}\right|
\end{gather}
of the equation~\eqref{vaneeva:eq_with_exact_solution_exp} with $\varepsilon=1$.


We continue the consideration by studying whether the equations from class~\eqref{vaneeva:class_vFH_exp}  
possessing nontrivial Lie symmetry properties, i.e. having the maximal Lie invariance algebras of dimension 
two or three, have nontrivial (i.e., inequivalent to Lie ones) regular reduction operators.
It has been already remarked that constant coefficient equations from class~\eqref{vaneeva:class_vFH_exp} do not admit 
such reduction operators \cite{vaneeva:Arrigo&Hill&Broadbridge1993,
vaneeva:Clarkson&Mansfield1993}.
Hence it is needless to consider cases~4 and~6 of table~1 as well as case~5 connected with case 6 by point transformation~\eqref{vaneeva:eq_add_eq_tr_exp}. 
As case 1 reduces to case 2 with the same transformation~\eqref{vaneeva:eq_add_eq_tr_exp}, we have to study only two cases, namely cases~2 and~3.
We substitute the pairs of values of the parameter-functions $F$ and $H$ corresponding to cases~2 and~3 
into system~\eqref{vaneeva:EqDet_exp} in order to find relevant values for~$b$. 
We ascertain that $b_t=0$ is a necessary condition for existing non-Lie regular reduction operators 
for equations with the above values of $(F,H)$. 
This is why we can use equations~\eqref{vaneeva:eq_FH_exp} instead of~\eqref{vaneeva:EqDet_exp} for further studying.

The investigation of case~3 of table~1 leads to the conclusion that there are no non-Lie regular reduction operators for this case.

The functions $F$ and $H$ presented in case 2 of table~1 satisfy~\eqref{vaneeva:eq_FH_exp} if and only if $\beta=2(1-\alpha)$, 
i.e. they have the form $F=\alpha x^{-1}$, $H=2(1-\alpha)x^{-2}$, and $k_1=k_2=0$. 
The corresponding value of~$b$ is $b=-(1+\alpha)x^{-1}$. 
Hence $\alpha\neq-1$ since otherwise $b=0$.
Substituting the derived form of the function~$b$ into the formulas~\eqref{vaneeva:eq_exact_solution_exp1} and~\eqref{vaneeva:eq_exact_solution_exp2}, 
we find that the equation
\begin{equation}\label{vaneeva:Eq_vFH_example5}
v_t=v_{xx}+\frac{\alpha}x v_x+\varepsilon e^v+\frac {2(1-\alpha)}{x^2}
\end{equation}
has the families of exact solutions
\[
v=-2\ln\left|\dfrac{\sqrt 2(1+\alpha)}{2s_1 x}\,{\cosh\left( s_1t+ \frac{s_1x^2}{2(1+\alpha)}+s_2\right)}\right|,
\]
if $\varepsilon=1$ and 
\begin{gather*}
v=-2\ln\left|\dfrac{\sqrt 2(1+\alpha)}{2s_1 x}\,{\sinh\left( s_1t+ \dfrac{s_1x^2}{2(1+\alpha)}+s_2\right)}\right|,\\[.5ex]
v=-2\ln\left|\dfrac{\sqrt 2(1+\alpha)}{2s_1 x}\,{\cos\left( s_1t+ \dfrac{s_1x^2}{2(1+\alpha)}+s_2\right)}\right|,\\[.5ex]
v=-2\ln\left|\dfrac{\sqrt{2}(1+\alpha)}{2\,x}\left(t+ \dfrac{x^2}{2(1+\alpha)}+c_2\right)\right|
\end{gather*}
if $\varepsilon=-1$. 
Recall that $s_1$, $s_2$ and~$c_2$ are arbitrary constants, $s_1\ne0$.

As a representative of the preimage of equation~\eqref{vaneeva:Eq_vFH_example5} with respect to the transformation~\eqref{vaneeva:gauge_exp}, 
we can choose the equation
\begin{gather}\label{vaneeva:Eq_preimaged_example5a}
x^\alpha u_t=\left(x^\alpha u_x\right)_x+\varepsilon x^{\alpha+2} e^u.
\end{gather}
Solutions of this equation can be easily constructed from the above solutions of equation~\eqref{vaneeva:Eq_vFH_example5} 
using the transformation $u=v-2\ln|x|$.
If $\alpha=1$, the chosen equation~\eqref{vaneeva:Eq_preimaged_example5a} can be replaced, e.g., by
$x u_t=\left(x u_x\right)_x+\varepsilon x e^{u}$ 
which is just another representation of equation~\eqref{vaneeva:Eq_vFH_example5}.

Non-Lie exact solutions of the equation
\begin{equation*}
v_t=v_{xx}+\left(\frac{\alpha}x+\mu x\right) v_x+\varepsilon e^v+\frac {2(1-\alpha)}{x^2}+2\mu,
\end{equation*}
where $\alpha\neq-1$ (case 1 of table~1), can be easily obtained from exact solutions of the equation~\eqref{vaneeva:Eq_preimaged_example5a}
using the transformation~\eqref{vaneeva:eq_add_eq_tr_exp}. 
The corresponding reduction operator has the form~\eqref{vaneeva:EqRedOp_exp} with $b=-(1+\alpha)x^{-1}-\mu x$.

We also prove the following assertions.

\begin{proposition}
Equations from class~\eqref{vaneeva:class_vFH_exp} with $F={\rm const}$ or $H={\rm const}$ 
may admit only nontrivial regular reduction operators that are equivalent to operators 
of the form~\eqref{vaneeva:EqRedOp_exp}, where the function $b$ does not depend on the variable $t$.
\end{proposition}

\begin{proposition}
Any reduction operator of an equation from class~\eqref{vaneeva:class_vFH_exp}, having the form~\eqref{vaneeva:EqRedOp_exp} with $b_{xx}=0$,
is equivalent to a Lie symmetry operator of this equation.
\end{proposition}

The proofs of these propositions are quite cumbersome and will be presented elsewhere. 

\subsection*{Acknowledgements}
The research of ROP was supported  by the Austrian Science Fund (FWF),  project P20632. 
OOV and ROP are grateful for the hospitality provided by the University of Cyprus.
ROP sincerely thanks the Cyprus Research Promotion Foundation (project number $\Pi$PO$\Sigma$E$\Lambda$KY$\Sigma$H/$\Pi$PONE/0308/01) 
for support of his participation in the Fifth Workshop on Group Analysis of Differential Equations and Integrable Systems.

{\footnotesize 

$^*$\phantom{$^*$}Available at http://www.imath.kiev.ua/$\sim$\,fushchych

\noindent
$^{**}$Available at http://www.imath.kiev.ua/$\sim$\,appmath/Collections/collection2006.pdf
}

\LastPageEnding
\end{document}